\documentclass[runningheads]{llncs}
\usepackage[T1]{fontenc}
\usepackage{bbding}
\usepackage{hyperref}
\hypersetup{colorlinks=true,
	linkcolor=blue,
	filecolor=blue,      
	urlcolor=blue,
	citecolor=blue}
    
\usepackage{graphicx}
\usepackage{booktabs}   % 绘制三线表
\usepackage{multirow}   % 合并单元格
\usepackage{tabularx}   % 自动调整列宽
\usepackage{subcaption}

\newcolumntype{Y}{>{\centering\arraybackslash}X}
\begin{document}
\title{Scaffolding Critical Engagement with GenAI: Transforming Ethnic Minority Preparatory Students’ Collaborative Discourse in Prompt Engineering Tasks}
\titlerunning{Scaffolding Ethnic Minority Students' Critical Engagement with GenAI}
\author{Deliang Wang\inst{1}\orcidID{0009-0008-6488-0234} \and
Cunling Bian\inst{2}$^($\Envelope$^)$\orcidID{0000-0003-0731-9228} }
%Third Author\inst{3}\orcidID{2222--3333-4444-5555}}
%
\authorrunning{D. Wang and C. Bian.}
\institute{Faculty of Education, The University of Hong Kong, Hong Kong, China\\
\email{wdeliang@connect.hku.hk} \and
Department of Education, Ocean University of China, Qingdao, China\\
\email{clbian@ouc.edu.cn}
} 
%Springer Heidelberg, Tiergartenstr. 17, 69121 Heidelberg, Germany
\maketitle              % typeset the header of the contribution
\begin{abstract}
Generative AI (GenAI) holds significant promise for advancing educational equity among ethnic minority students by broadening access to learning resources and mitigating linguistic barriers. However, these benefits are counterbalanced by the risk of cognitive laziness, whereby students may treat GenAI as an answer engine or shortcut rather than as a partner in thinking. This design-based research investigated how pedagogical scaffolding can shift students from passive consumption to critical co-creation with GenAI. The study involved 78 ethnic minority preparatory students in China participating in a three-week GenAI course that integrated a human-in-the-loop workflow and teacher modeling with contrasting cases to disrupt uncritical reliance on GenAI. We employed epistemic network analysis to examine collaborative discourse, thematic analysis to analyze student reflections, and paired-samples \textit{t}-tests to assess changes in prompt self-efficacy. Results revealed a phenomenon of strategic repurposing: initially, students instrumentalized strategy talk to coordinate efficient copying; however, after the intervention, they realigned strategic planning to scaffold critical evaluation and peer co-construction. Qualitative findings further indicated that the teacher’s scaffolding helped students overcome their initial authority bias and prompt paralysis, repositioning themselves as active gatekeepers of AI-generated content; these shifts were corroborated by a significant increase in students’ prompt self-efficacy. The study suggests that, particularly for ethnic minority students, technical training alone is insufficient; educators should design targeted pedagogical interventions around human--AI collaboration to prevent cognitive complacency and cultivate epistemic agency.

\end{abstract}
\section{Introduction}
Prior to the advent of generative AI (GenAI), educational inequality was largely defined by the unequal distribution of resources \cite{yang2014analysis}. Students from high socioeconomic status (SES) families typically enjoyed privileged access to high-quality educational materials and personalised tutoring \cite{huang2010parental}, whereas their low-SES peers depended primarily on limited school-based provision \cite{tranter2012unequal}. This disparity is further exacerbated for ethnic minority students from under-resourced regions. Research indicates that these populations in China face a double disadvantage: significantly lower household income compared to the Han majority \cite{chia2023educational,Gustafsson2003ethnic} and substantial linguistic barriers \cite{postiglione2015education}. For many minority students, Mandarin (Putonghua) is a second language, and limited proficiency often impedes their performance in high-stakes examinations such as the Gaokao (i.e., unversity entrance examinations), placing them in a structurally underprivileged position even after they enter higher education \cite{tsung2015minority,postiglione2015education}.

%The emergence of GenAI provides a promising solution to these issues, enabling democratic access to high-quality, personalized educational support regardless of a student's geographic,  economic, and ethinic background \cite{james2024levelling}. Theoretically, a minority student in a remote area can now access the same AI tutor as a student in the most developed areras (e.g., Beijing and Shanghai). However, scholars warn that the digital divide is shifting from a access divide to a usage divide \cite{hendawy2024intensified}. While the access gap narrows, the usage gap may widen. Developed areas, such as Beijing and Shenzhen, have launched projects on providing students with systematic training in digital and AI literacy \cite{BMEC2025}, enabling them to use GenAI as a tool for critical inquiry. In contrast, disadvantaged students, lacking such pedagogical guidance, are prone to cognitive reliance/laziness \cite{fan2025beware} -- using GenAI merely as an answer engine or a shortcut to bypass cognitive effort rather than as a collaborative partner.

The emergence of GenAI has been considered a promising solution to mitigate these forms of inequality by enabling more democratic access to high-quality, personalised educational support, regardless of a student's geographic, economic, or ethnic background \cite{james2024levelling}. Theoretically, a minority student in a remote region can now access the same AI tutor as a student in highly developed areas such as Beijing or Shanghai. However, recent scholarship cautions that the digital divide is shifting from an \emph{access divide} to a \emph{usage divide} \cite{hendawy2024intensified}. While disparities in access may be narrowing, disparities in the quality and sophistication of use may be widening. Developed regions, such as Beijing and Shenzhen, have launched large-scale initiatives to provide students with systematic training in digital and AI literacy \cite{BMEC2025}, enabling them to employ GenAI as a tool for critical inquiry and knowledge construction. In contrast, students in disadvantaged contexts, who lack such pedagogical support, are more vulnerable to cognitive reliance or laziness \cite{fan2025beware}, using GenAI primarily as an answer engine or shortcut to bypass cognitive work rather than as a collaborative partner in thinking.

%This risk is particularly acute for students in the ethnic minority preparatory program -- a specialized affirmative action policy in China. This program allows minority students from under-resourced regions to enter elite universities with lower entrance scores, requiring them to complete a one-year bridging curriculum before beginning their undergraduate majors \cite{liu2023ethnic}. While GenAI offers these linguistically disadvantaged students access to educational resources \cite{james2024levelling}, it simultaneously presents a trap of cognitive offloading. Without pedagogical guidance, students may succumb to cognitive reliance, using GenAI as a shortcut rather than a tool for inquiry \cite{fan2025beware}.Consequently, bridging the digital divide for this specific population requires more than mere access; it demands empirical evidence on how to effectively transform their interaction with GenAI from passive consumption to critical co-creation through targeted pedagogical interventions.

This risk is particularly salient for students enrolled in the ethnic minority preparatory program, a specialised affirmative action policy in China. This program admits minority students from under-resourced regions to elite universities with lower entrance scores, on the condition that they complete a one-year bridging curriculum prior to beginning their undergraduate majors \cite{liu2023ethnic}. While GenAI can, in principle, provide these linguistically disadvantaged students with unprecedented access to learning resources \cite{james2024levelling}, it simultaneously creates a risk of cognitive laziness. Without carefully designed pedagogical guidance, students may become dependent on GenAI as a shortcut, rather than using it as a tool for exploration and inquiry \cite{fan2025beware}. Consequently, bridging the digital divide for this specific population requires more than providing technological access; it calls for empirical evidence on how to transform their interaction with GenAI from passive consumption to critical co-creation through targeted pedagogical interventions.

%To address this imperative, it is essential to first understand the students' unguided interaction patterns and then examine the impact of scaffolding. Therefore, the present study tracks the evolution of ethnic minority preparatory students' collaborative discourse throughout a three-week GenAI course. Specifically, we capture their initial engagement in a natural state to identify authentic challenges, and subsequently implement teacher scaffolding to facilitate a shift in their cognitive engagement. By analyzing the trajectory of their discourse across these distinct phases, this study aims to reveal the mechanism of this transformation. The research questions (RQs) guiding this study are as follows:

To address this imperative, it is first necessary to understand how students interact with GenAI in the absence of explicit guidance and then examine how scaffolding reshapes these interaction patterns. The present study therefore tracks the evolution of ethnic minority preparatory students' collaborative discourse over the duration of a three-week GenAI-intensive course. Specifically, we first examine their initial engagement with GenAI under naturalistic, minimally guided conditions to identify authentic challenges. We then introduce teacher scaffolding aimed at fostering more reflective and critical engagement with GenAI. By analysing the trajectory of students' discourse across these distinct phases, the study seeks to illuminate the mechanisms underlying the transformation from dependency to more active epistemic agency. The research questions (RQs) guiding this study are as follows:

\begin{itemize}
    \item \textbf{RQ1}: How does students' collaborative discourse evolve from the initial unguided phase to the post-scaffolding phase?
    \item \textbf{RQ2}: How do students perceive the challenges of initial collaboration and the role of teacher scaffolding in shifting their engagement with GenAI?
    \item \textbf{RQ3}: Is the shift towards critical collaboration associated with improvements in students' prompt self-efficacy?
\end{itemize}
%Specifically, we initially invited these students to use GenAI to complete a collaborative task without any intervention. Subsequently, we organized a workshop on critically collaborate with GenAI for them and they were required to perform another taught students.

\section{Related Work}
\subsection{Educational dilemmas of ethnic minority students in China}
\label{subsection:2.1}
%China is a unified multi-ethnic country composed of the Han majority (approximately 91.1\% of the population) and 55 recognized ethnic minority groups, such as the Uyghur, Tibetan, and Hui, who predominantly reside in peripheral regions, including western and border areas \cite{national2021main}. While these groups possess distinct cultural and linguistic heritages, they often face structural disadvantages in the mainstream education system. As highlighted in the introduction, these students frequently come from under-resourced regions where economic capital is relatively scarce \cite{hannum2006geography}. However, the dilemma extends beyond mere resource access to profound linguistic and pedagogical disconnects.

China is a unified multi-ethnic state comprising a Han majority (approximately 91.1\% of the population) and 55 officially recognized ethnic minority groups, such as the Uyghur, Tibetan, and Hui, who predominantly reside in peripheral regions, including western and border areas \cite{national2021main}. While these groups possess rich and distinct cultural and linguistic heritages, they often face structural disadvantages within the mainstream education system. As highlighted in the introduction, many ethnic minority students come from under-resourced regions where economic capital is relatively scarce \cite{hannum2006geography}. However, the dilemma extends beyond mere resource access to profound linguistic and pedagogical disconnects.

%On one hand, for many minority students, Mandarin (Putonghua) is a second language. Transitioning from a mother tongue to Mandarin-medium instruction in K-12 and higher education imposes a heavy extraneous cognitive load \cite{rong2009development,zhang2023multilingual}. Research indicates that the continuous effort required for linguistic transcoding competes with the cognitive resources needed for higher-order thinking, making deep comprehension and critical argumentation significantly more taxing for these learners than for their Han peers \cite{tsung2015minority}.

On the one hand, for many minority students, Mandarin (Putonghua) functions as a second language. The transition from learning in their mother tongue to Mandarin-medium instruction in K–12 and higher education imposes a substantial extraneous cognitive load \cite{rong2009development,zhang2023multilingual}. Research indicates that the continuous effort required for linguistic transcoding competes with the cognitive resources needed for higher-order thinking, making deep comprehension and critical argumentation considerably more demanding for these learners than for their Han peers \cite{tsung2015minority}.

%On the other hand, research indicates that in secondary education within ethnic minority regions, pedagogical practices often exhibit a stronger inclination towards traditional, teacher-centered models that prioritize textbook memorization for preparing high-stakes standardized examinations (e.g., Gaokao), rather than student-centered approaches fostering critical thinking and inquiry-based learning \cite{postiglione2015education}. Consequently, these students may develop passive epistemic beliefs, viewing learning as the reception of static facts rather than the dynamic construction of knowledge \cite{chan2004relational}. This ingrained habit of passivity likely pose a significant challenge for them when they enter university environments \cite{tsung2015minority} and workplaces where 21st-century skills are emphasized, such as critical thinking and authentic problem solving competencies.

On the other hand, studies show that secondary education in many ethnic minority regions is often dominated by traditional, teacher-centred pedagogies that prioritize textbook memorization and exam preparation for high-stakes assessments (e.g., the Gaokao), rather than student-centred approaches that cultivate critical thinking and inquiry-based learning \cite{postiglione2015education}. As a result, these students may develop passive epistemic beliefs, viewing learning primarily as the reception of static facts rather than the active construction of knowledge \cite{chan2004relational}. This ingrained habit of passivity likely poses a significant challenge when they enter university environments \cite{tsung2015minority} and, later, workplaces where 21st-century skills—such as critical thinking and authentic problem-solving competencies—are emphasised.

\subsection{The benefits and cognitive risks of GenAI in education}
Recent empirical studies and meta-analyses have provided robust evidence for the positive impact of GenAI on student learning outcomes. Comprehensive reviews indicate that GenAI interventions yield moderate to large effect sizes on academic performance, significantly enhancing learning efficiency and achievement across various disciplines \cite{gu2025effects,gokouglu2025effects}. For example, GenAI has been found to positively influences students’ lower-order cognitive skills (e.g., basic understanding) and identified as a promising approach for cultivating students’ higher-order cognitive skills \cite{xia2025impact}. In collaborative contexts, such as digital storytelling, GenAI tools have proven effective in augmenting team creativity performance, serving as an ideation partner that helps students overcome creative blocks and generate high-quality content \cite{wei2025effects}.

These performance gains, however, may be accompanied by unintended cognitive costs. Scholars caution that although GenAI can improve the quality of students’ products, it may inadvertently encourage cognitive laziness or cognitive dependency \cite{fan2025beware}. Fan et al. \cite{fan2025beware} found that students working with GenAI reported significantly lower mental effort and reduced metacognitive monitoring compared with those using conventional tools, suggesting a tendency to offload cognitive work onto the AI. Likewise, Han et al. \cite{han2025identifying} observed that students—particularly those with a high need for cognitive closure—were inclined to accept AI-generated answers prematurely to avoid uncertainty, rather than engaging in critical evaluation. These findings point to a paradox: students may produce ostensibly better outputs while engaging less deeply with the underlying content, as GenAI replaces rather than productively scaffolds the rigorous cognitive processes required for deep understanding \cite{fan2025beware,bhullar2024chatgpt}.

%This risk of cognitive offloading is particularly concerning for the ethnic minority preparatory students described in Section \ref{subsection:2.1}. Given their existing linguistic barriers and potentially ingrained passive epistemic beliefs, these students may be even more susceptible to treating GenAI as an authoritative answer engine rather than a collaborative peer. If general university populations struggle with metacognitive laziness \cite{fan2025beware}, the double disadvantage faced by ethnic minority students likely amplifies this vulnerability. Therefore, simply providing access to GenAI is insufficient; it is imperative to implement pedagogical scaffolding that explicitly fosters critical AI literacy, guiding students to shift from passive acceptance to critical co-creation \cite{gu2025effects,ouaazki2024generative}.

This risk of cognitive offloading is especially concerning for the ethnic minority preparatory students. Given their existing linguistic barriers and potentially entrenched passive epistemic beliefs, these learners may be particularly prone to treating GenAI as an authoritative answer engine rather than a dialogic partner in inquiry. If mainstream university students already struggle with cognitive laziness in GenAI-supported learning \cite{fan2025beware}, the ``double disadvantage" experienced by ethnic minority students is likely to magnify this vulnerability. Consequently, merely providing access to GenAI is inadequate; there is a pressing need for carefully designed pedagogical scaffolding that cultivates critical AI literacy and deliberately supports a shift from passive acceptance of AI outputs to reflective, critical co-creation \cite{gu2025effects,ouaazki2024generative}.

\section{Method}
\subsection{Context and participants}

%This study was conducted within an intact class of the ethnic minority preparatory program at an elite university in northern China. As a key affirmative action policy, this program admits students from under-resourced ethnic minority regions to foster educational equity and ethnic solidarity. These students are required to complete a one-year bridging curriculum designed to strengthen their academic foundations before matriculating into their undergraduate majors.

%These students were invited to participate a three-week course on GenAI, where basic knowledge of GenAI and prompt engineering skills were taught. Students were required to conduct collaborative prompting engineering tasks. Thus, they were assigned into 13 groups and each group was composed of six ethnic minority students. 

This study was conducted with an intact class (N=78) enrolled in an ethnic minority preparatory programme at an elite university in northern China. As a key affirmative action initiative, this program admits students from under-resourced ethnic minority regions with the aim of promoting educational equity and ethnic solidarity. Admitted students are required to complete a one-year bridging curriculum designed to strengthen their academic foundations before matriculating into their undergraduate majors.

All students in this class were invited to participate in a three-week course on GenAI, in which basic concepts of GenAI and prompt engineering were introduced. Students were required to complete collaborative prompt-engineering tasks. To this end, they were assigned to 13 groups, with each group comprising six ethnic minority students.

\begin{figure}
    \centering
    \includegraphics[width=\linewidth]{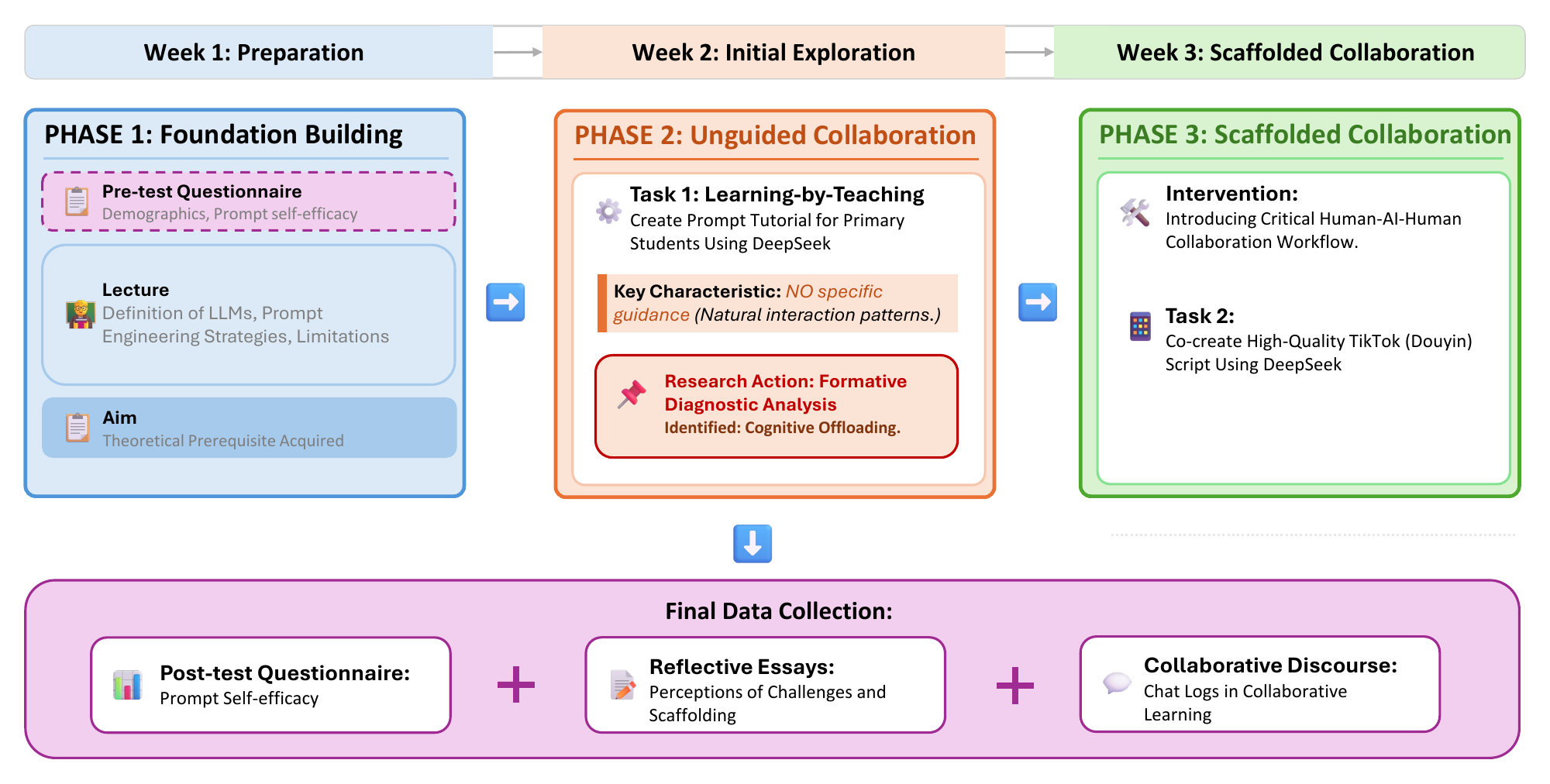}
    \caption{Experimental procedures.}
    \label{fig:procedures}
\end{figure}

\subsection{Experimental procedures and intervention}
%The study employed a design-based research approach spanning three weeks, structured into three distinct phases: baseline preparation, initial unguided exploration, and scaffolded co-creation. Each phase was implemented through a weekly two-hour workshop.
As Figure \ref{fig:procedures} shows, the study adopted a design-based research approach spanning three weeks, structured into three phases: preparation, initial unguided exploration, and scaffolded collaboration. Each phase was implemented through a weekly two-hour workshop.

%The primary objective of the first week was to establish a foundational understanding of GenAI among participants. To ascertain the students' entry-level characteristics, they were first administered a pre-test questionnaire collecting demographic data, prior GenAI experience, and initial self-efficacy in prompt engineering. Following this, the instructor delivered a comprehensive lecture covering the definition of LLMs, the strategies of prompt engineering, potential technical limitations such as hallucinations, and ethical considerations, and practical use cases of LLMs. This phase ensured that all students possessed the necessary theoretical prerequisites to operate DeepSeek for subsequent collaborative tasks.
The primary objective of Week~1 was to establish a foundational understanding of GenAI among participants. To ascertain students' entry-level characteristics, a pre-test questionnaire was first administered to collect demographic information, prior GenAI experience, and initial self-efficacy in prompt engineering. Subsequently, the instructor delivered a lecture covering the definition of LLMs, core strategies of prompt engineering, potential technical limitations such as hallucinations, ethical considerations, and illustrative use cases of LLMs. This phase ensured that all students possessed the theoretical prerequisites necessary to operate DeepSeek in the subsequent collaborative tasks.

%In the second week, students engaged in their first collaborative task without specific pedagogical intervention regarding collaborative strategies. Groups were tasked with creating a prompt engineering tutorial for primary school students using DeepSeek. This learning-by-teaching task design was selected to cognitively challenge students to externalize their understanding of prompting strategies. Crucially, this phase served as a diagnostic window to capture the students' authentic, unguided interaction with GenAI and peers. After the second week, we quickly reviewed the collaborative discourse generated and performed a formative diagnostic analysis. We found several groups frequently copied DeepSeek's outputs with minimal discussion or critical verification.
In Week~2, students undertook their first collaborative task without any explicit pedagogical intervention regarding collaborative strategies. Each group was tasked with using DeepSeek to create a prompt-engineering tutorial for primary school students. This learning-by-teaching task was designed to cognitively challenge students to externalize and organize their understanding of prompting strategies. At the same time, this phase served as a diagnostic window to capture students' authentic, unguided interaction with GenAI and with peers. Immediately after Week~2, the research team reviewed the collaborative discourse and conducted a formative diagnostic analysis. This analysis revealed that several groups frequently copied DeepSeek's outputs with minimal discussion or critical verification.

\begin{figure}
    \centering
    \includegraphics[width=\linewidth]{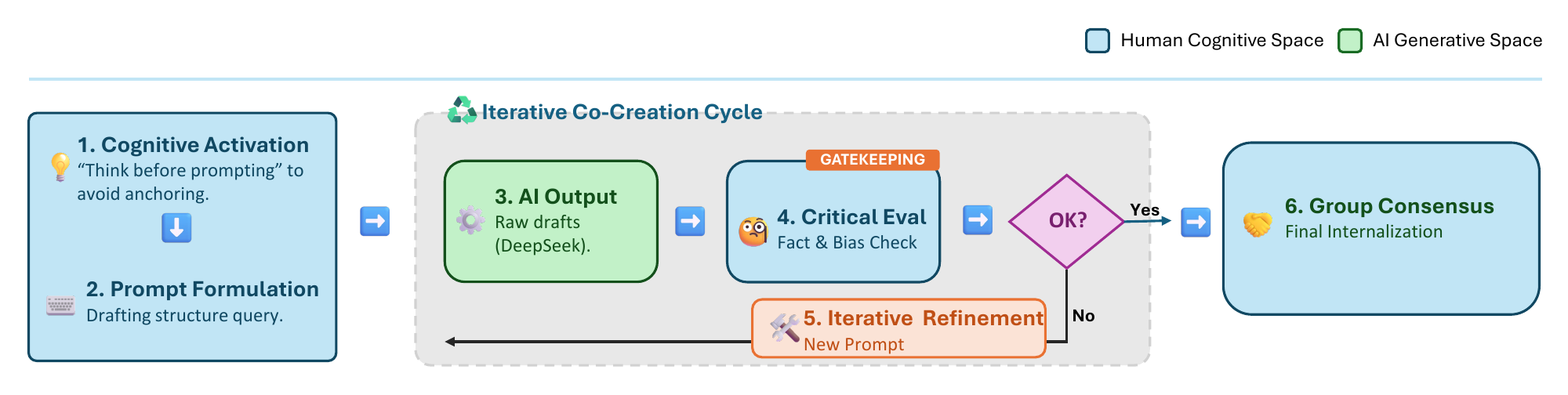}
    \caption{The human-in-the-loop workflow.}
    \label{fig:workflow}
\end{figure}
%To address the passive engagement identified in Week 2, an intervention scaffolding  critical human-AI-human collaboration  was designed and implemented in Week 3. First, the instructor introduced a structured human-in-the-loop workflow to students. Drawing on the principle of cognitive activation, students were instructed to formulate their own initial ideas before querying the AI. This step was explicitly designed to mitigate the tendency to offload all cognitive effort to the AI \cite{fan2025beware}. to prevent anchoring effects [1]. Furthermore, to foster critical evaluative literacy, the workflow required groups to treat AI outputs as raw materials necessitating fact-checking and bias scrutiny rather than authoritative answers \cite{xia2025impact}. The strategy also emphasized iterative refinement, encouraging students to use follow-up prompts to interrogate and refine the AI's output \cite{wei2025effects}.
To address the passive engagement identified in Week~2, an intervention scaffolding critical human–AI–human collaboration was designed and implemented in Week~3. First, the instructor introduced a structured human-in-the-loop workflow, as shown in Figure \ref{fig:workflow}. Drawing on the principle of cognitive activation, students were instructed to generate their own initial ideas before querying the AI. This step was explicitly intended to mitigate tendencies to offload cognitive effort to the AI \cite{fan2025beware} and to reduce anchoring effects. In addition, to foster critical evaluative literacy, the workflow required groups to treat AI outputs as provisional materials requiring fact-checking and scrutiny for potential bias, rather than as authoritative answers \cite{xia2025impact}. The workflow further emphasized iterative refinement, encouraging students to employ follow-up prompts to interrogate, adjust, and improve the AI's output \cite{wei2025effects}.

%Second, to facilitate the internalization of these strategies, the intervention utilized modeling with contrasting cases \cite{schwartz2011practicing}. The instructor presented two simulated collaboration transcripts: a negative example exhibiting passive pasting and superficial agreement, and a positive example demonstrating critical discussion and iterative inquiry. By comparing these contrasting cases, students were guided to explicitly identify and model desirable collaborative behaviors. Following this intervention, groups undertook the second authentic task: using DeepSeek to co-create a high-quality script for a specialized TikTok  video. This task required higher creativity and contextual judgment, serving as a testing ground for the newly acquired critical strategies.
Second, to facilitate the internalization of these strategies, the intervention employed instructor modelling with contrasting cases \cite{schwartz2011practicing}. The instructor presented two simulated collaboration transcripts: a negative example illustrating passive copying and superficial agreement, and a positive example demonstrating critical discussion and iterative inquiry. By comparing these contrasting cases, students were guided to explicitly identify and emulate productive collaborative behaviours. Following this intervention, groups undertook a second authentic task: using DeepSeek to co-create a high-quality script for a specialised TikTok video. This task demanded greater creativity and contextual judgement, thereby providing a testing ground for the newly acquired critical collaboration strategies.

%At the conclusion of Week 3, data collection was finalized. Students completed a post-test questionnaire assessing their prompt efficacy and submitted reflective essays detailing their perceived challenges during the initial collaboration and their perceptions of how the teacher’s scaffolding influenced their engagement with GenAI.
At the conclusion of Week~3, data collection was completed. Students filled out a post-test questionnaire assessing their prompt-engineering self-efficacy and submitted reflective essays describing the challenges they encountered during the initial collaboration, as well as their perceptions of how the instructor's scaffolding influenced their engagement with GenAI.

\subsection{Data collection and analysis}
\subsubsection{Collaborative discourse data}
\paragraph{Coding scheme}
%To investigate how collaborative discourse evolve from the initial unguided phase to post-scaffolding phase (RQ1), we used a coding scheme to annotate students' engagement through group chat logs and then analyzed the annotations using the epistemic network analysis (ENA) method.
To investigate how collaborative discourse evolved from the initial unguided phase to the post-scaffolding phase (RQ1), we applied a coding scheme to annotate students' engagement in group chat logs and then analyzed these annotations using epistemic network analysis (ENA).

%The coding scheme, as seen in Table \ref{tab:coding_scheme}, was devised based on the ICAP framework\cite{chi2014icap} and adapted for the specific context of human-GenAI collaborative learning \cite{fan2025beware,wei2025effects,lukevsova2026clue}. It categorizes each message into five hierarchical levels: off-topic, passive, active, constructive, and interactive. 
The coding scheme, presented in Table~\ref{tab:coding_scheme}, was developed based on the ICAP framework \cite{chi2014icap,wang2025investigating} and adapted to the specific context of human–GenAI collaborative learning \cite{fan2025beware,wei2025effects,lukevsova2026clue}. It classifies each message into five hierarchical levels of engagement: off-topic, passive, active, constructive, and interactive.

%Messages unrelated to the learning task were coded as off-Topic (OT). \textit{Passive} engagement is characterized by mechanical information sharing (P.Share) or uncritical acceptance (P.Acpt). \textit{Activ}e involves activating prior knowledge to manipulate information, such as summarizing AI responses (A.Rprt) or asking operational questions (A.Qst), but without generating new inferences. \textit{Constructive} represents the core of critical engagement, where learners generate new ideas beyond the immediate information presented. This level encompasses three distinct cognitive behaviors: articulating abstract strategies for prompt engineering (C.Stra), critically evaluating the quality, style, or validity of specific AI outputs (C.Crit), and iteratively refining prompts based on reflection (C-Refi). Finally, \textit{Interactive} involves dialogic co-construction where students build upon (I.Build) or negotiate (I.Neg) peers' ideas regarding the AI content, rather than interacting solely with the tool. Following the ``rule of high-road coding", the highest level of engagement exhibited in a message was assigned as the final code.
Messages unrelated to the learning task were coded as off-topic (OT). \textit{Passive} engagement is characterized by mechanical information sharing (P.Share) or uncritical acceptance (P.Acpt). \textit{Active} engagement involves the activation of prior knowledge to manipulate or reorganize information, such as summarizing AI responses (A.Rprt) or asking operational questions (A.Qst), but without generating new inferences. \textit{Constructive} engagement represents the core of critical engagement, where learners generate ideas that go beyond the information directly provided. This level includes three types of cognitive behavior: articulating abstract strategies for prompt engineering (C.Stra), critically evaluating the quality, style, or validity of AI outputs (C.Crit), and iteratively refining prompts based on reflection (C.Refi). Finally, \textit{Interactive} engagement involves dialogic co-construction, where students build on (I.Build) or negotiate (I.Neg) peers' ideas in relation to the AI-generated content, rather than interacting solely with the tool. Following the ``high-road'' coding rule, the highest level of engagement evidenced in a message was assigned as its final code.

\begin{table}[h]
    \centering
    \scriptsize % 保持小字体
    \caption{The coding scheme for analyzing collaborative discourse.}
    \label{tab:coding_scheme}
    
    % 将行间距设为 1.1 (紧凑但依然可读)，如果想要极限紧凑可改为 1.0
    \renewcommand{\arraystretch}{1.1} 
    
    % 列宽定义：
    % 第1列 1.5cm (极限压缩)
    % 第2列 2.8cm (刚好放下 Code 名字)
    % 第3、4列自动分配剩余空间
    \begin{tabularx}{\textwidth}{@{} 
        >{\raggedright\arraybackslash}p{1.5cm} 
        >{\raggedright\arraybackslash}p{2.8cm} 
        >{\raggedright\arraybackslash}X 
        >{\raggedright\arraybackslash}X 
    @{}}
        \hline
        \textbf{Level} & \textbf{Code} & \textbf{Definition} & \textbf{Examples} \\
        \hline
        
        % Level 0
        Level 0: Off-Topic & OT (Off-Topic) & 
        Content unrelated to the task (e.g., gossip, lunch, socializing). & 
        ``Are we going to the canteen later?'' \\ \hline
        % 删除了 \addlinespace 以减少间隔
        
        % Level 1
        % \multirow{行数}{宽度}{内容}
        % 注意：如果右边内容换行很多，这里的行数可能需要手动微调（比如写 3 或 4）来居中
        \multirow{4}{=}{Level 1: Passive} 
        & P.Share (Mechanical Sharing) & 
        Copy-pasting AI text, sharing screenshots, or sending links without analysis. & 
        [Sends a screenshot of DeepSeek] / ``Here is the answer.'' \\
        
        & P.Acpt (Uncritical Acceptance) & 
        Blind agreement with AI or peers. Simple acknowledgments or stickers. & 
        ``Okay.'' / ``Received.'' / [Sticker: Nodding] \\ \hline
        
        % Level 2
        \multirow{4}{=}{Level 2: Active} 
        & A.Rprt (Active Reporting) & 
        Summarizing, paraphrasing, or translating the AI's output. & 
        ``DeepSeek suggested three scenes: a classroom, a playground, and a dorm.'' \\
        
        & A.Qst (Operational Questioning) & 
        Questions focused on task logistics or tool operation rather than content depth. & 
        ``How do I input this prompt?'' / ``Did you send the screenshot yet?'' \\ \hline
        
        % Level 3
        \multirow{8}{=}{Level 3: Constructive} 
        & C.Stra (Strategy Talk) & 
        Articulating abstract rules or strategies for prompting engineering. & 
        ``To get a good script, we must specify the target audience clearly in the prompt.'' \\ 
        
        & C.Refi (Iterative Refinement) & 
        Proposing specific changes to the prompt or refining the output based on reflection. & 
        ``Let's change the prompt to `act as a funny blogger' to get a better hook.'' \\
        
         & C.Crit (Critical Evaluation) & 
        Explicitly judging the quality, style, or validity of the AI's output. & 
        ``This script is too formal; it doesn't sound like a Douyin video at all.'' \\\hline
        
        % Level 4
        \multirow{4}{=}{Level 4: Interactive} 
        & I.Build (Co-Construction) & 
        Extending, elaborating, or challenging a peer's contribution. & 
        ``I agree with your idea to make the tone funnier, and we can ask DeepSeek to use internet slang.'' \\
        
        & I.Neg (Negotiation) & 
        Discussing conflicting views among group members or negotiating choices. & 
        ``You think we should use Script A, but I think Script B is better. Let's vote.'' \\
        
        \hline
    \end{tabularx}
\end{table}

\paragraph{Coding procedure}
%To ensure the reliability of the coding, two researchers first independently coded a random subsample of the dataset (approx. 10\% of the total messages). Any discrepancies were resolved through discussion to refine the coding manual. Then the two researchers independently coded the remaining dataset. Inter-rater reliability was calculated using Cohen's Kappa, yielding a coefficient of $\kappa = [0.88]$ (e.g., 0.85), indicating substantial agreement. All discrepancies were resolved through discussions.
To ensure the reliability of the coding, two researchers first independently coded a randomly selected subsample of the dataset (approximately 20\% of all messages). Discrepancies were discussed and resolved, leading to refinement of the coding manual. The two researchers then independently coded the remaining messages. Inter-rater reliability, calculated using Cohen's kappa, yielded a coefficient of $\kappa = 0.88$, indicating substantial agreement. All remaining disagreements were resolved through discussion.

\paragraph{ENA} 
%ENA is a  quantitative ethnography method used to analyze the co-occurrences of coded data in educational dialogues \cite{shaffer2017epistemic}. It can visualize connections and strength of associations among different discourse elements within a context using a network and compare the difference of networks between different contexts \cite{wang2024tutors,wang2025more}. When using ENA to analyze collaborative discourse, we established four key components: codes, conversations, the unit of analysis, and the stanza. Codes represented the specific engagement codes that each message was annotated with. Conversations indicated the talk exchanged within a specific activity, referring to the message between students during a collaborative task in our study. The unit of analysis was defined as individuals, ideas, organizations, or any other entity whose structure of connections that need to be modeled, referring to the combination of Week (i.e., Week 2 or Week) and Group (i.e., which group). Lastly, the stanza referred to the range of co-occurrence between codes modeled by ENA, with a moving stanza window of 5 set for this study. 
ENA is a quantitative ethnographic method used to analyze patterns of co-occurrence among coded elements in discourse data \cite{shaffer2017epistemic}. It models and visualizes the structure and strength of connections among discourse elements as networks, and allows for comparison of these networks across different conditions \cite{wang2024tutors,wang2025more,gao2024discourse}.

In applying ENA to the collaborative discourse in this study, we specified four key components: codes, conversations, units of analysis, and stanzas. \textit{Codes} corresponded to the engagement codes assigned to each message using the aforementioned scheme. \textit{Conversations} referred to the messages exchanged among students within a given collaborative task. The \textit{unit of analysis} was defined as the combination of week (Week~2 or Week~3) and group, such that each unit represented the discourse produced by a particular group in a given week. Finally, the \textit{stanza} specified the window of co-occurrence for ENA modeling; in this study, we employed a moving stanza window of five consecutive messages.
\subsubsection{Reflective data}
%Students' reflective essays were analyzed using thematic analysis \cite{braun2006using}. The first author first repeatedly read the data and marked the data with initial codes related to the challenge of initial collaboration and the role of teacher scaffolding teacher in shifting their engagement with GenAI. Next, themes were identified inductively from the initial codes, and then refined and given names. To ensure credibility, the second author reviewed the coding scheme in order to increase its content validity. Revisions were made until the two authors agreed on the scheme. Finally, the first and second authors applied the revised coding scheme to the data separately.  All of the discrepancies then underwent further discussion, and a final agreement was reached after modification, as appropriate
Students' reflective essays were analyzed using thematic analysis \cite{braun2006using}. The first author repeatedly read the essays and generated initial codes related to the challenges experienced during the initial collaboration and the perceived role of teacher scaffolding in reshaping students’ engagement with GenAI. Next, themes were inductively identified from these initial codes, refined, and clearly named. To ensure credibility, the second author reviewed the coding scheme to enhance its content validity. Revisions were made until both authors reached agreement on the scheme. Finally, the first and second authors independently applied the revised coding scheme to the full dataset. All remaining discrepancies were discussed and resolved.

\subsubsection{Prompt self-efficacy data}
%The prompt self-efficacy data was collected from a adapted scale \cite{mun2025exploring} designed to assess users’ confidence in their ability to craft effective prompts for LLMs. All eight items are scored on a 5-point Likert scale ranging from 1 (“Strongly disagree”) to 5 (“Strongly agree”). The paired-samples t test was used to compare whether there was a significant improvement in students' prompt self-efficacy after the course.
Prompt self-efficacy data were collected using an adapted scale \cite{mun2025exploring} designed to assess users’ confidence in their ability to craft effective prompts for LLMs. The scale comprises eight items rated on a 5-point Likert scale ranging from 1 (“Strongly disagree”) to 5 (“Strongly agree”). A paired-samples \textit{t}-test was conducted to examine whether there was a significant improvement in students' prompt self-efficacy after the course.
\section{Results}
\subsection{Collaborative discourse}
\begin{figure}[h]
    \centering
    % --- 第一张图 (左边) ---
    \begin{subfigure}[b]{0.55\textwidth}
        \centering
        % 插入图片，宽度设为子图宽度的100%
        \includegraphics[width=\linewidth]{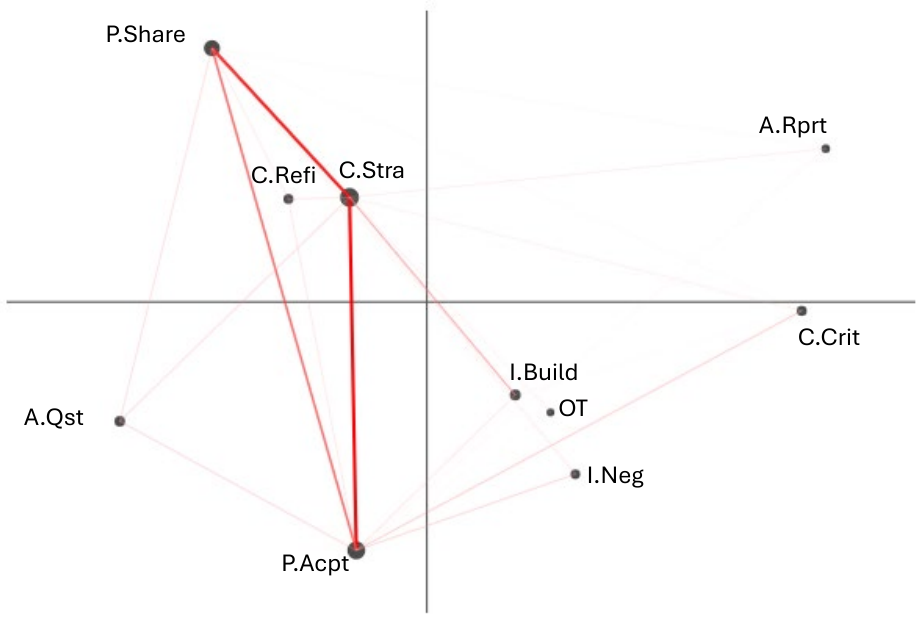} 
        \caption{The overall network of cognitive engagement in unguided collaboration.}
        \label{fig:left}
    \end{subfigure}
    \\
    %\hfill % 这个指令非常关键，它将两个子图推向两侧
    % --- 第二张图 (右边) ---
    \begin{subfigure}[b]{0.55\textwidth}
        \centering
        \includegraphics[width=\linewidth]{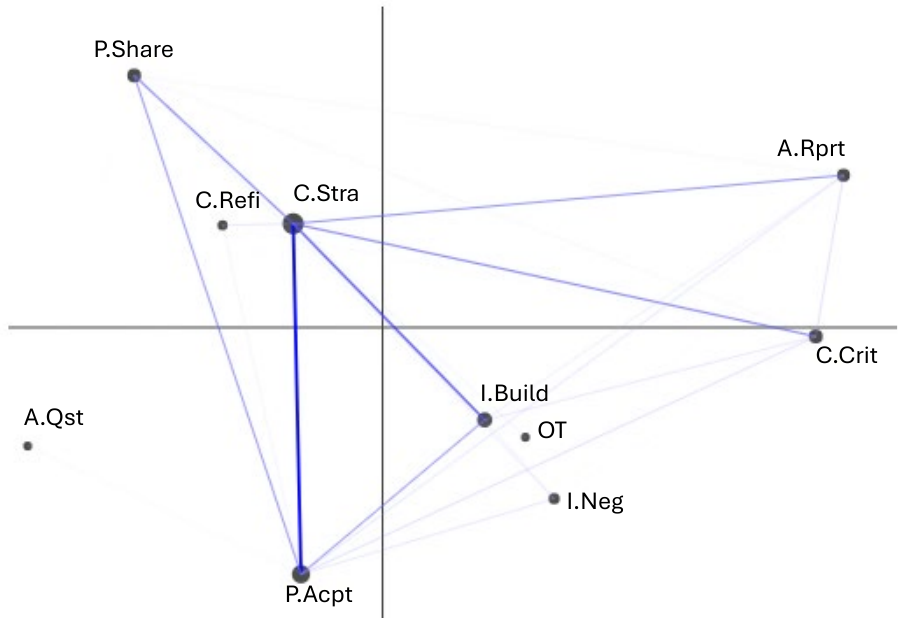}
        \caption{The overall network of cognitive engagement in scaffolded collaboration.}
        \label{fig:right}
    \end{subfigure}
    
    \caption{Students' epistemic network.}
    \label{fig:both}
\end{figure}

%Figure \ref{fig:both} shows the visualized networks of students' cognitive engagement in unguided collaboration (Figure \ref{fig:left}) and scaffolded collaboration (Figure \ref{fig:right}). In these figures, codes were represented as nodes, with their size indicating the prevalence. Larger nodes corresponded to more frequent occurrences of the cognitive behavior. The connections between codes were depicted as edges, with the thickness of an edge representing the frequency of co-occurrence between the linked codes. Thicker edges indicated a higher rate of co-occurrence, indicating a stronger association between the respective cognitive behavior.
Figure \ref{fig:both} visualizes the epistemic networks of students' cognitive engagement during the unguided (Figure \ref{fig:left}) and scaffolded (Figure \ref{fig:right}) phases. In these sociograms, nodes represent cognitive codes, with their size proportional to the frequency of occurrence, while edges represent the relative strength of co-occurrence between behaviors. Thicker edges denote a higher probability that two cognitive behaviors appeared in close proximity, revealing the underlying structural patterns of student discourse.
%\vspace{-5mm}

%Based on the node size in Figure \ref{fig:left},  students during the unguided collaboration frequently engaged in behavior of P.Share, P.Accept, and C.Stra. By contrast, Figure \ref{fig:right} indicates that, in addition to the three shared behavior (i.e., P.Share, P.Accept, and C.Stra), they frequently engaged in behavior of A.Rprt, C.Crit, and I.Build during the scaffolded collaboration, suggesting a shift toward critical evaluation and collaborative construction.
As illustrated in Figure \ref{fig:left}, the unguided discourse network reveals a paradoxical pattern regarding the role of strategic planning. While students frequently engaged in Strategy Talk (C.Stra), the strongest co-occurrences linked this high-level code directly to passive sharing (P.Share) and uncritical acceptance (P.Acpt). This counter-intuitive structural configuration suggests that without scaffolding, students’ strategic discussions were instrumental rather than epistemic—aimed at efficiently generating content for immediate cognitive offloading rather than deep inquiry. The prominent connections between these nodes indicate a plan-and-paste loop, where strategy formulation served primarily to expedite the completion of the task via mechanical copying.

%Figure \ref{fig:difference} illustrates the subtracted network of students' cognitive engagement between the unguided and scaffolded collaboration. It displays a graph by subtracting nodes and connection weights from two overall networks (i.e., comparing the unguided collaboration network and scaffolded collaboration network). The dominance of the red color in the subtracted network indicates that the unguided collaboration had a greater number of co-occurrences of the specified codes compared to the scaffolded collaboration, and vice versa. 
In stark contrast, the scaffolded network (Figure \ref{fig:right}) demonstrates a fundamental topological shift towards high-order engagement. While P.Share and P.Acpt persist, the network structure has reconfigured around constructive and interactive nodes. Crucially, the connectivity of C.Stra transformed significantly; it formed a strong coherent cluster with Summarizing (A.Rprt) and Critical Evaluation (C.Crit), as well as linking to Co-construction (I.Build). This indicates that the intervention successfully repurposed strategy talk: instead of serving as a prelude to passive copying, strategies were deployed to guide the summarization and critical scrutiny of AI outputs. The emergence of the C.Stra—A.Rprt—C.Crit triad suggests a more sophisticated workflow where students articulated a strategy, summarized the AI's response to check alignment, and then critically evaluated its quality.

\begin{figure}[h]
    \centering
    \includegraphics[width=0.75\linewidth]{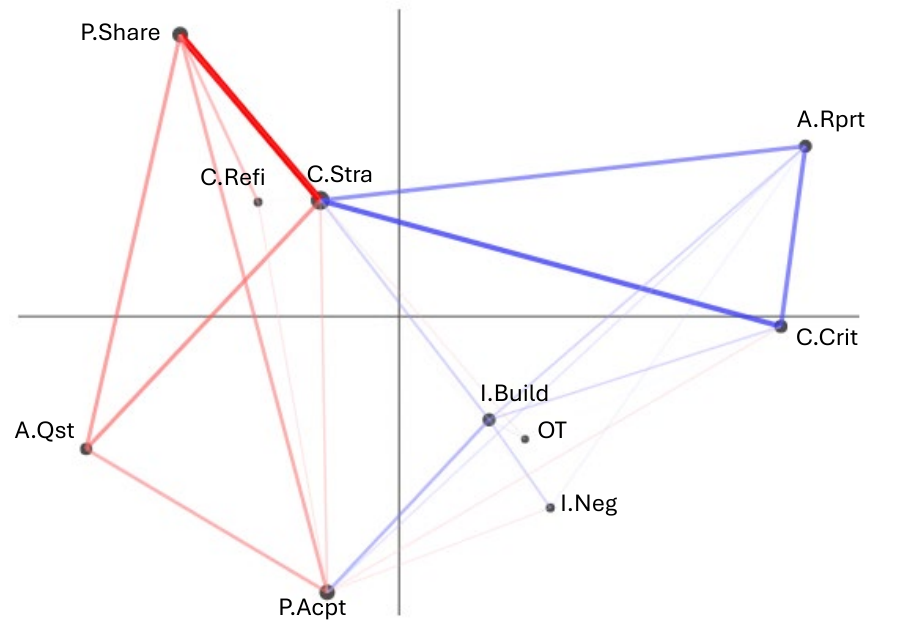}
    \caption{The subtracted network showing differences in students' cognitive engagement between \textcolor{red}{unguided collaboration} and \textcolor{blue}{scaffolded collaboration}.}
    \label{fig:difference}
\end{figure}

%Clearly, we can see that students during the unguided collaboration showed more patterns related to P.Share, P.Acpt, A. Qst, and C.Stra.  By contrast, students during the scaffolded collaboration demonstrated more behavior patterns about P.Acpt, A.Rprt, C.Crit, C.Stra, and I.Build. The differences suggest that students at the unguided stage were more inclined to passively copy and paste GenAI answers (P.Share) or ask operational questions (A.Qst) while students after teacher scaffolding more tended to summarize (A.Rprt) and critically evaluate GenAI answers (C.Crit), as well as co-construct with their peers (I.Build). The findings partially suggest the effectiveness of the intervention in fostering students' critical engagement with GenAI.
Figure \ref{fig:difference} explicitly quantifies these structural transformations through the subtracted network. The dominance of red edges (representing associations stronger in the unguided phase) connecting P.Share and C.Stra visually highlights the ``metacognitive disconnect" characteristic of the initial phase, where planning collapsed into passive execution. Conversely, the dense blue connections (representing associations stronger in the scaffolded phase) linking C.Stra, C.Crit, A. Rprt, and I.Build provide robust empirical evidence of the intervention's efficacy. These patterns suggest that teacher scaffolding did not merely increase the frequency of constructive behaviors but fundamentally altered the functional role of strategic planning, fostering a transition from passive consumption to critical co-creation with GenAI.
%\vspace{-5mm}
\subsection{Students' perceptions of challenges and scaffolding}
The thematic analysis of students' reflections indicated that their initial passive reliance on GenAI arose from a combination of limited prompt literacy and a pronounced epistemic asymmetry between themselves and the tool. With respect to the challenges of unguided collaboration, students frequently reported a sense of ``prompting paralysis'': they did not simply lack operational knowledge of the tool, but struggled to translate vague, abstract intentions into explicit prompts that the LLM could process effectively. Several participants described the frustration of ``knowing what to ask in mind but being unable to phrase it in a way the AI understood'', which often resulted in irrelevant or only partially useful responses that they felt unable to revise or redirect.

More fundamentally, the analysis suggested that such cognitive offloading (e.g., direct copy-pasting of AI-generated text) was driven by an authority bias compounded by low academic self-efficacy. Students explicitly acknowledged that they tended to accept AI outputs because they perceived them as ``more professional and organized'' than their own writing, or because, under time pressure, the AI's responses appeared ``good enough.'' In this sense, many students positioned GenAI as an epistemic authority to be deferred to rather than a tool to be interrogated and appropriated, which in turn created a psychological barrier to critical engagement.

Students' accounts of the scaffolding phase indicated that the teacher intervention—particularly the modeling with contrasting cases—functioned as an important turning point that reconfigured their collaborative norms with GenAI. They did not experience the scaffolding merely as technical training in prompt design, but as a deliberate challenge to their initial blind trust in algorithmic outputs. The juxtaposition of high- and low-quality collaboration examples helped students recognize their own passive practices (e.g., unconscious acceptance) as detrimental to learning. In response, many reported a shift in their perceived agency: they began to see themselves as active evaluators or gatekeepers of AI-generated content rather than passive recipients.

Following the intervention, several groups described establishing explicit norms to ``cross-check possible AI hallucinations'' and to engage in ``iterative negotiation'' with the tool by refining prompts and requesting revisions instead of accepting the first response. As one student reflected, the scaffolding encouraged them ``not to accept the AI's answer directly, but to examine its logic and adapt it to our specific context.'' Overall, these reflections suggest that the pedagogical guidance helped re-centre human judgement and disciplinary knowledge as the primary drivers of the human–GenAI collaborative process.

\subsection{Prompt self-efficacy}
%Table \ref{tab:ttest} presents the results of the paired-samples t test that compare students' prompt self-efficacy before and after the teacher intervention. Although 78 students participated in this study, only 71 students completed both the pre- and post-test questionnaire. The results indicate that in the pre-test, students reported a mean score of 3.757 (SD = 0.654) in their prompt self-efficacy, while in the post-test, they reported an average score of 3.991 (SD = 0.589). The t test (t = 3.729, p < .001) indicates that students' perceived prompt self-efficacy in the post-test is significantly higher that that in the pre-test, suggesting students' improvement in prompt self-efficacy.
Table~\ref{tab:ttest}  reports the results of the paired-samples \textit{t}-test comparing students' prompt self-efficacy before and after the teacher intervention. Although 78 students participated in the study, complete pre- and post-test data were available for 71 students. The results show that, in the pre-test, students reported a mean prompt self-efficacy score of 3.757 ($SD = 0.654$), whereas in the post-test, the mean score increased to 3.991 ($SD = 0.589$). The paired-samples \textit{t}-test ($t = 3.729$, $p < .001$) indicates that students’ perceived prompt self-efficacy was significantly higher after the intervention than before it, suggesting a meaningful improvement in their confidence in crafting prompts.
%\vspace{-5mm}
\begin{table}[htbp]
    \centering
    \caption{The results of the paired-samples t test.}
    \label{tab:ttest}
    
    % {Y Y Y Y Y Y} 表示6列全部使用我们定义的自动等宽居中列
    % \linewidth 表示表格总宽度等于当前页面的版心宽度
    \begin{tabularx}{\linewidth}{Y Y Y Y Y Y}
        \hline
        
        % 表头
        & \textit{N} & \textit{Mean} & \textit{SD} & \textit{t} & \textit{p} \\
        
        \hline
        
        % 第一行：pre-test
        % 注意：第一列如果是文字，居中可能不好看。
        % 如果想让第一列左对齐但保持等宽，可以将第一个 Y 改为 >{\raggedright\arraybackslash}X
        pre-test  & 71 & 3.757 & 0.654 & \multirow{2}{*}{3.729} & \multirow{2}{*}{$<$ .001} \\
        
        % 第二行：post-test
        post-test & 71 & 3.991 & 0.589 & & \\
        
        \hline
    \end{tabularx}
\end{table}
%\vspace{-5mm}
\section{Discussion and Conclusion}
This study addresses the need to support underrepresented ethnic minority students in learning how to collaborate critically with GenAI. While GenAI can provide a powerful linguistic scaffold for these learners, our findings offer empirical evidence that access alone does not guarantee deep learning; without pedagogical guidance, it may inadvertently foster cognitive complacency. We conducted a three-week design-based research intervention with 78 ethnic minority preparatory students and examined (a) the evolution of collaborative discourse using epistemic network analysis, (b) students' perceptions using thematic analysis, and (c) changes in prompt self-efficacy using statistical analysis. With respect to discourse evolution, the study revealed a trajectory from instrumental to epistemic engagement: initially, students used GenAI primarily to enable passive copying, whereas after the intervention they increasingly engaged in critical evaluation and peer co-construction.  The thematic analysis of students' reflective essays further indicated that their initial passive reliance on GenAI stemmed from a combination of limited prompt literacy and a pronounced epistemic asymmetry between themselves and the tool, whereas the teacher's scaffolding helped them overcome their initial authority bias and prompt paralysis and reposition themselves as active gatekeepers of AI outputs. This transition was corroborated by a significant increase in students' prompt self-efficacy. Overall, the findings underscore the necessity of metacognitive scaffolding for learners from diverse or under-resourced backgrounds. For ethnic minority students in particular, we argue that technical training alone is insufficient; educators should design targeted pedagogical interventions around human--AI collaboration to avoid cognitive complacency and cultivate epistemic agency.

Despite these contributions, several limitations should be acknowledged. First, the study was conducted within a specific context—an ethnic minority preparatory program in China—so the findings may not readily generalize to other populations or educational settings. Second, the three-week duration, although sufficient to observe immediate shifts in discourse and self-efficacy, does not guarantee the long-term maintenance of these critical collaborative habits. Future research should employ longitudinal designs to examine whether norms of critical human--AI collaboration persist as students transition into their undergraduate studies, and how these interaction patterns adapt to different disciplinary tasks and demands.

\bibliographystyle{splncs04}
\bibliography{mybibliography}

\end{document}